\documentclass[preprint,12pt,nofootinbib]{revtex4-1}

\usepackage[brazil, english]{babel}
\usepackage{graphicx}
\usepackage{epsfig}
\usepackage{amssymb}
\usepackage{amsmath}
\usepackage{ulem}
\usepackage[T1]{fontenc} % if needed
\usepackage[utf8]{inputenc}
\usepackage{color}

\usepackage[titletoc]{appendix}

\begin{document}

\title{Chiral Symmetry Breaking and Restoration in
(2+1) Dimensions from Holography:
Magnetic and Inverse Magnetic Catalysis}
 
 \author{Diego M. Rodrigues$^{1} $}
\email[Eletronic address: ]{diegomr@if.ufrj.br}
\author{Danning Li$^{2} $}
\email[Eletronic address: ]{lidn@itp.ac.cn}
\author{Eduardo Folco Capossoli$^{1,3}$}
\email[Eletronic address: ]{educapossoli@if.ufrj.br}
\author{Henrique Boschi-Filho$^{1,}$}
\email[Eletronic address: ]{boschi@if.ufrj.br}  
\affiliation{$^1$Instituto de F\'\i sica, Universidade Federal do Rio de Janeiro, 21.941-972 - Rio de Janeiro-RJ - Brazil \\
 $^2$Department of Physics and Siyuan Laboratory, Jinan University, Guangzhou 510632, China \\$^3$Departamento de F\'\i sica and Mestrado Profissional em Pr\'aticas da Educa\c c\~ao B\'asica (MPPEB), Col\'egio Pedro II, 20.921-903 - Rio de Janeiro-RJ - Brazil}

\begin{abstract}
We study the chiral symmetry breaking and restoration in $ (2+1) $-dimensional  gauge theories from the holographic hard and softwall models. We describe the behavior of the chiral condensate in the presence of an external magnetic field for both models at finite temperature. For the hardwall model we find Magnetic Catalysis (MC) in different setups. For the softwall model we find Inverse Magnetic Catalysis (IMC) and MC in different situations. We also find for the softwall model a crossover transition from IMC to MC at a pseudocritical magnetic field. This study also shows spontaneous symmetry breaking for both models. Interestingly, for $B=0$ in the softwall model we found a nontrivial expectation value for the chiral condensate.
\end{abstract}

%\keywords{Holographic Models, (2+1) Gauge Theories, (Inverse) Magnetic Catalysis.}

\maketitle

\newcommand{\limit}[3]
{\ensuremath{\lim_{#1 \rightarrow #2} #3}}

\section{Introduction}   
Much theoretical progress on the nonperturbative physics of relativistic quantum field theories (RQFT) has been achieved recently. In particular, the role played by an external magnetic field in these RQFT has been investigated in many works, especially in connection with QCD \cite{Klimenko:1990rh, Klimenko:1991he, Gusynin:1994re,  Miransky:2002rp,Gatto:2010pt, Shovkovy:2012zn,Miransky:2015ava, Preis:2010cq, Dudal:2015wfn, Mamo:2015dea, Li:2016gfn, Evans:2016jzo}. In the context of lattice QCD, for the chiral phase transition, an inverse magnetic catalysis (IMC), has been observed, i.e., the decreasing of the critical temperature $(T_c)$ with increasing magnetic field $(B)$ for $eB \sim 1\, $GeV$^2$ \cite{Bali:2011qj} and  more recently for $eB \sim 3\, $GeV$^2$ \cite{Endrodi:2015oba}. This is in contrast with what would be expected: a magnetic catalysis (MC), meaning the increasing of the critical temperature with increasing magnetic field \cite{Shovkovy:2012zn}.

A promising approach to investigate the phenomena of IMC and MC is based on the AdS/CFT correspondence, or Holographic Duality \cite{Maldacena:1997re, Gubser:1998bc, Witten:1998qj, Witten:1998zw}. Such duality has become very useful to address strongly coupled gauge theories, including the nonperturbative regime of QCD. However, in order to reproduce the nonperturbative physics of QCD, one must break the conformal invariance in the original AdS/CFT correspondence. There are two well-known models which do this symmetry breaking: the hardwall \cite{Polchinski:2001tt, Polchinski:2002jw, BoschiFilho:2002ta, BoschiFilho:2002vd, BoschiFilho:2005yh,  Capossoli:2013kb, Rodrigues:2016cdb} and softwall \cite{Karch:2006pv, Colangelo:2007pt,FolcoCapossoli:2016ejd} models. In particular, some researches have used these models to discuss the IMC as a perturbation in powers of the magnetic field \cite{Dudal:2015wfn, Mamo:2015dea, Li:2016gfn, Evans:2016jzo}, and can only be trusted for weak fields $eB < 1 $GeV$^2$. Since their approach is perturbative in the magnetic field $B$, they could not predict what would happen for $eB>1 $GeV$^2$.

Concerning free fermions in ($ 2+1 $) dimensions in the presence of a magnetic field we know that, perturbatively, the phenomenon of magnetic catalysis is due to the effect of the magnetic field itself, which is a strong catalyst of dynamical chiral symmetry breaking, leading to the generation of a nonzero chiral condensate in the chiral limit ($m \rightarrow0$) given by \cite{Gusynin:1994re}:
\begin{equation}\label{cond}
\left\langle \bar{\psi}\psi \right\rangle \propto  \dfrac{eB}{2\pi}.
\end{equation}
At finite temperature, also in the perturbative regime, it was shown in \cite{Das:1995bn} that the chiral condensate is extremely unstable, meaning that it vanishes as soon as we introduce a heat bath with and without chemical potential.

In this work we investigate, nonperturbatively, the chiral phase transition and symmetry restoration in $ (2+1) $-dimensional holographic gauge theories in an external magnetic field from the hard and softwall models. We confirm the picture mentioned before, but we also find IMC, in the softwall model, for small magnetic fields (in units of the string tension squared), in agreement with the lattice results in $ (3+1) $ dimensions, while the traditional MC behavior happens for large magnetic fields in this nonperturbative approach. In the hardwall model we only find MC. At finite temperature our approach predicts a very fast decrease in the chiral condensate as we increase the temperature, which means that we do not observe the instability found in the perturbative thermal field theory computation presented in \cite{Das:1995bn}.

\section{The Chiral Condensate and Chiral Symmetry Breaking in the Holographic Framework}

\subsection{Background Geometry}
Via Kaluza-Klein dimensional reduction, the supergravity theory on $ AdS_4\times S^7 $ may be consistently truncated to Einstein-Maxwell Theory on $ AdS_4$ \cite{Herzog:2007ij}. The action for this theory, in Euclidean signature, is given by
\begin{equation} \label{AdS4Action}
S = -\dfrac{1}{2\kappa^2_4}\int d^{4}x \sqrt{g}\left(\mathcal{R} -2\Lambda - L^{2}F_{MN}F^{MN}\right),
\end{equation}
where $ \kappa^2_4 $ is the 4-dimensional coupling constant, which is proportional to the 4-dimensional Newton's constant $( \kappa^2_4\equiv8\pi G_4 )$, $ d^{4}x\equiv d\tau\,dx_{1}\,dx_{2}\,dz $, $ \mathcal{R} $ is the Ricci scalar and $ \Lambda $ is the negative cosmological constant. 

The field equations coming from the bulk action \eqref{AdS4Action} together with the Bianchi identity are \cite{Herzog:2007ij}
\begin{eqnarray}
R_{MN} &=& 2L^2\left( F_{M}^{P}F_{NP}-\dfrac{1}{4}g_{MN}F^2\right)  - \dfrac{3}{L^2}g_{MN}, \label{FieldEquations}\\
\nabla_{M}F^{MN} &=& 0.
\end{eqnarray}
Our Ansatz for the metric and the background magnetic field to solve \eqref{FieldEquations} are given by 
\begin{eqnarray}
ds^2 &=& \dfrac{L^2}{z^2}\left( f(z)d\tau^2 + \dfrac{dz^2}{f(z)} + dx^2_1 + dx^2_2\right), \label{AnsatzMetrica} \\
F &=& B\, dx_{1}\wedge dx_{2} \label{ansatzB}.
\end{eqnarray}
Using this Ansatz the field equations \eqref{FieldEquations} are simplified and given by
\begin{eqnarray} 
z^2f''(z)-4zf'(z)+6f(z)-2B^{2}z^4-6 = 0, \label{FieldequationSimplified1} \\
zf'(z)-3f(z)-B^{2}z^4+3 = 0, \label{FieldequationSimplified2}
\end{eqnarray} 
with $ F^2 $ in \eqref{AdS4Action} given by
\begin{equation} \label{MagneticField}
F^2 = \dfrac{2B^{2}z^4}{L^4}.
\end{equation}

The general solution for Eqs. \eqref{FieldequationSimplified1} and \eqref{FieldequationSimplified2} is given by:
\begin{equation} \label{Gen_Sol}
f(z)=1+B^2z^4 + bz^3.
\end{equation}
If we impose the horizon condition $f(z=z_H)=0$ one finds:
\begin{eqnarray}
f_{BT}(z) &=& 1 + B^{2}z^3(z-z_H) - \dfrac{z^3}{z^{3}_H}\,. \label{fBT}
\end{eqnarray} 
This solution was found recently in \cite{Rodrigues:2017iqi, Rodrigues:2017cha}. The Eq. \eqref{fBT}, corresponds to the $ AdS_4 $ in the presence of a background magnetic field at finite temperature, $ T $, where $ z_H $ is the horizon position, such that $ f_{BT}(z=z_H)=0 $.  One can note that this solution indeed satisfies both differential equations \eqref{FieldequationSimplified1} and \eqref{FieldequationSimplified2}. Note that the solution Eq. \eqref{fBT} implies the existence of an inner and outer horizon. The outer horizon satisfies $f_{BT}'(z=z_H) < 0$ and is the physically relevant one. 
The temperature is given by the Hawking formula
\begin{equation} \label{T1}
T = \dfrac{|f_{BT}'(z=z_H)|}{4\pi}.
\end{equation} 
Using \eqref{fBT} and the condition $f_{BT}'(z=z_H) < 0$ we have
\begin{equation}\label{T2}
T(z_{H},B) = \dfrac{1}{4\pi}\left( \dfrac{3}{z_{H}} - B^2 z_{H}^3 \right),  \quad z_{H}^4< \dfrac 3 {B^2}. 
\end{equation}

In what follows and in the rest of this work we set the $ AdS $ radius $ L=1 $.

\subsection{Holographic Setup for Chiral Symmetry Breaking}

Here we describe how to realize the chiral symmetry breaking of $ SU(N_f)_{L}\times SU(N_f)_{R}\rightarrow SU(N_f)_{\mathrm{diag}}  $ in the hardwall and softwall models. For both models we consider the action: 
\begin{equation}\label{ChiralAction}
S = -\dfrac{1}{2\kappa^2_4}\int d^{3}x \, dz \sqrt{g}\,e^{-\Phi}\mathrm{Tr}\left(D_{M}X^{\dagger}\,D^{M}X + V_{X} -(F_{L}^2+F_{R}^2) \right), 
\end{equation}
where $ X $ is a complex scalar field, $ D_{M} $ is the covariant derivative defined as $ D_{M}X = \partial_{M}X + iA_{M}^{L}X - iXA_{M}^{R} $, with $ A_{M}^{L,R} $ being the chiral left- and right-handed gauge fields, $ F_{MN} $ the field strength defined as $ F_{MN} = \partial_{M}A_{N} - \partial_{N}A_{M} - i[A_{M},A_{N}] $ and $ V_{X} = M_{4}^2X^{\dagger}X + \lambda(X^{\dagger}X)^2+... $ is the potential for the complex scalar field, where $ M_{4} $ is the mass of the complex scalar field $ X $. From the AdS$ _4 $/CFT$ _{3} $ correspondence we have
\begin{equation}
M_{4}^2 = \Delta(\Delta-3).
\end{equation}
Since the complex scalar field in 4 spacetime dimensions is supposed to be dual to the chiral condensate $ \sigma\equiv\left\langle \bar{\psi}\psi\right\rangle $ in 3 spacetime dimensions, whose dimension is $ \Delta=2 $, we have therefore $ M_{4}^2=-2 $. So, we have the potential $ V_{X} = -2X^2 + \lambda X^4 $, where the first term is just the mass term and the second is the term needed to realize the spontaneous symmetry-breaking mechanism \cite{Gherghetta:2009ac,Chelabi:2015cwn,Chelabi:2015gpc}.

The field equations coming from \eqref{ChiralAction} are given by
\begin{equation}\label{ChiralFieldEquations}
D_M\left[ \sqrt{g} \;  e^{-\Phi(z)} g^{MN} D_N {X}\right] - \sqrt{g} 
e^{-\Phi(z)}\partial_{X}V_X = 0.
\end{equation}
Here, we also assume that the expectation value for X takes a diagonal form $ \left\langle X \right\rangle = \frac 12 \, {\chi}\, I_2 $ for the $ SU(2) $ case \cite{Dudal:2015wfn,Li:2016gfn}, where $ I_2 $ is the $ 2\times2 $ identity matrix. For the general case of $ SU(N_f) $ we would have $ \left\langle X \right\rangle  = \frac{1}{\sqrt{2N_f}}\, \chi\, I_{N_f} $, where the factor $ \sqrt{2N_f} $ is introduced to maintain the kinetic term of $ \chi $ canonical \cite{Chelabi:2015cwn,Chelabi:2015gpc}. In addition, we assume that $ \chi(x^{\mu},z)\equiv\chi(z) $. With all these assumptions we can write \eqref{ChiralFieldEquations} as
\begin{equation}\label{ChiralFieldEquations2}
\chi''(z) + \left(-\frac{2}{z} - \Phi'(z) + \frac{f'(z)}{f(z)}\right) \chi'(z) - \dfrac{1}{z^2 f(z)}\partial_{\chi}V(\chi) = 0,
\end{equation}
where $ ' $ means derivative with respect to $ z $. For our purposes the dilatonic field is $ \Phi(z)=kz^2 $ with $k$ a dimensional constant. For the hardwall model we set $k=0$. On the other hand for the softwall model $k \not= 0$. Finally, we take $ V(\chi)\equiv\mathrm{Tr}\,V_X = -\chi^2 + \lambda\chi^4$, in both models. 

First, let us consider only the quadratic term in the potential $V(\chi)$, i.e., $V(\chi) = -\chi^2$, which implies a linear and analytically solvable equation of motion. In this case we can clearly see that the UV behavior ($ z\rightarrow0 $) of \eqref{ChiralFieldEquations2} takes the form
\begin{equation}\label{ChiralFieldEquations3}
\chi''(z) -\frac{2}{z}\chi'(z) + \dfrac{2}{z^2}\chi(z) = 0,
\end{equation}
whose solution is given by
\begin{equation}\label{UV behavior of X}
\chi(z\rightarrow0) = c_{1}z + c_{2}z^2,
\end{equation}
where $ c_1 $ and $ c_2 $ are integration constants. Since the complex scalar field $ \chi(z) $ is dual to the operator $\left\langle \bar{\psi}\psi\right\rangle $, we can identify these two integration constants as proportional to: 
\begin{eqnarray}
c_1 & \propto & m_{q}\,\\
c_2 & \propto & {\sigma}\,,
\end{eqnarray}
with $ m_{q} $ being the fermion mass in $ (2+1) $ dimensions, and $ \sigma=\left\langle \bar{\psi}\psi\right\rangle $ is the chiral condensate.

In this work we solve numerically Eq. \eqref{ChiralFieldEquations2}  using the quartic potential $ V(\chi) = -\chi^2+\lambda\chi^4 $, with $ \lambda=1 $. The boundary conditions used to solve \eqref{ChiralFieldEquations2} are the UV ($ z\rightarrow0 $) behavior of $ \chi(z) $, that is, $\chi(z) = m_{q}z + \sigma z^2$ and the regularity of $ \chi(z) $ at the horizon, $ z_{H} $, meaning $ \chi(z_{H})<\infty $. 

\subsection{Zero temperature case}

Here the background metric is pure AdS, without singularity, and there is no natural boundary condition at the horizon. Numerically, one cannot take the horizon exactly at infinity. Thus, to analyze the chiral symmetry breaking in the confining region ($ T=0 $) we need to perform some coordinate transformations in order to bring the boundary condition at $z\rightarrow\infty$ to a finite position. To this end, we are going to split the analysis into two cases: (i) $ T = 0 $ and $ B = 0 $; (ii) $ T = 0 $ and $ B\neq0 $. 

In the first case (i) $ f(z) = 1 $ in Eq. \eqref{ChiralFieldEquations2}, so we have
\begin{equation}\label{ChiralFieldEquations4}
\chi''(z) + \left(-\frac{2}{z} - \Phi'(z) \right) \chi'(z) - \dfrac{1}{z^2}\left(-2\chi(z) + 4\lambda\chi^3(z)\right) = 0.
\end{equation}
In the $ z $ coordinate the two boundaries are $z\rightarrow0$ and $z\rightarrow\infty$. 

For the hardwall model, $ k=0 $, we can do the following coordinate transformation
\begin{equation}
z = \dfrac{1-t}{t}\,; \quad t = \dfrac{1}{1+z}.
\end{equation}
Under this transformation the two boundaries in the $ t $ coordinate become $ t = 1\, (z\rightarrow0) $ and $ t = 0\, (z\rightarrow\infty) $, and the equation of motion \eqref{ChiralFieldEquations4} becomes
\begin{equation}
\chi''(t) + \dfrac{2(2-t)}{t(1-t)}\chi'(t) + \dfrac{2\chi(t) -4\lambda\chi^3(t)}{t^2(1-t)^2} = 0,
\end{equation}
which has the following IR ($ t=0 $) and UV ($ t=1 $) asymptotic behaviors
\begin{eqnarray}
\chi_{_{IR}}(t) &=& \dfrac{1}{\sqrt{2\lambda}} + c\,t + \mathcal{O}(t^2), \\
\chi_{_{UV}}(t) &=& - m_{q} (t-1) + (m_{q} + \sigma)(t-1)^2 + \mathcal{O}((t-1)^3),
\end{eqnarray}
where $c$ is a constant and $ m_q $ and $ \sigma $ were defined in $ z $ coordinate by $\chi(z\rightarrow0) \equiv \chi_{_{UV}} = m_{q}z + \sigma z^2$. 

Now for the softwall model, still in case (i), we have $ \Phi(z) = - z^2 $, where we have set $ k = - 1 $. In this case equation \eqref{ChiralFieldEquations4} becomes
\begin{equation}\label{ChiralFieldEquations5}
\chi''(z) - \left(\frac{2}{z} - 2z \right) \chi'(z) - \dfrac{1}{z^2}(-2\chi(z) + 4\lambda\chi^3(z)) = 0,
\end{equation}
with the leading expansion at $ z\rightarrow\infty $ given by
\begin{equation}
\chi_{_{IR}}(z) = \dfrac{1}{\sqrt{2\lambda}} - c\,\dfrac{e^{-z^2}}{z}.
\end{equation}
while at $ z\rightarrow0 $ we still have $\chi(z\rightarrow0) \equiv \chi_{_{UV}} = m_{q}z + \sigma z^2$. Note that in this case (the softwall with $T=0$ and $B=0$) there is no need to change the coordinates to obtain the asymptotic behaviours at IR and UV due to the presence of the dilaton. 

Finally, in case (ii) for $ T = 0 $ and $B\neq0$, we have \eqref{ChiralFieldEquations2} with $ f(z) $ given by \eqref{fBT} with $z_{H} = \frac{3^{1/3}}{\sqrt{B}} $, where we have used the definition of the Hawking temperature \eqref{T2}. Doing the coordinate transformation $ s = z/z_H $, the boundaries in the $ s $ coordinate are $ s = 0 $ (UV), and $ s = 1 $ (IR), and the equation \eqref{ChiralFieldEquations2} in this coordinate becomes
\begin{equation} \label{ChiralFieldEquations6}
\chi''(s) - \left(\dfrac{2}{s} - \dfrac{2\sqrt{3}\,k}{B}s + \dfrac{12s^2}{(1-s)(1+2s+3s^2)}\right)\chi'(s) + \dfrac{-2\chi(s) + 4\lambda\chi^3(s)}{s^{2}(1-s)^{2}(1+2s+3s^2)} = 0.
\end{equation}
One can check that the near boundary ($ s\rightarrow 0 $) solution of the above equation is
\begin{equation}
\chi_{_{UV}}(s) = a\,s + b\,s^2,
\end{equation}
with the two integral constants $a$ and $b$ related to the fermion mass $ m_q $ and the chiral condensate $ \sigma $ by the following relations
\begin{equation}
a = {3^{1/4}} \dfrac{m_q }{\sqrt{B}}\,; \quad b = {3}^{1/2} \dfrac{\sigma}{B}.
\end{equation} 
Now, keeping the divergent terms as $s\rightarrow1$ in equation \eqref{ChiralFieldEquations6}, one reaches
\begin{equation}
\chi''(s) - \dfrac{2}{1-s}\chi'(s) + \dfrac{-2\chi(s) + 4\lambda\chi^3(s)}{6(1-s)^{2}} = 0, 
\end{equation}
whose solution is given by
\begin{equation}
\chi_{_{IR}}(s) = \dfrac{1}{\sqrt{2\lambda}} + c\,(1-s)^{-\frac{1}{2} + \frac{\sqrt{33}}{6}} + c' \times (\mathrm{singular\; solution})\,,
\end{equation}
with $ c $ and $c'$ being two integral constants at IR ($ s=1 $). In order to have a nonsingular solution we set $c'=0$. 

In the following sections we present our results for IMC and MC in the context of the chiral phase transition for the hardwall and softwall models. 

\section{Results for the Hardwall model}

In this section we present our results for the hardwall model concerning the dependence of the chiral condensate, $\sigma$, with the external magnetic field, $B$, for $T=0$ and some finite temperatures. We also study the dependence of $\sigma$ with temperature, $T$, for some values of magnetic field. These analyses allow us to investigate whether there is IMC or MC and the role played by the thermal effects. We will also investigate whether the chiral symmetry is spontaneous or explicitly broken at the nonpertubative level. It is worth to remember that we are working in (2+1) dimensions so that all observable quantities like $\sigma$, $B$, $T$, and $m_{q}$ will always be measured in units of the string tension ($\sqrt{\sigma_s}$). Then, the temperature and mass will be measured in units of $\sqrt{\sigma_s}$, for instance, in refs. \cite{Teper:1998te, Meyer:2003wx, Athenodorou:2016ebg}. On the other hand, the magnetic field and the chiral condensate will be measured in units of the string tension squared $(\sqrt{\sigma_{s}})^2$. One should note that the hardwall model can be defined setting $k= 0$ in Eq. \eqref{ChiralAction} for the dilaton field, given by $\Phi(z)=kz^2$ and introducing a hard cutoff $z_{max}$ such that the holographic coordinate $z$ obeys $0\leq z \leq z_{max}$ \cite{Polchinski:2001tt, Polchinski:2002jw, BoschiFilho:2002ta, BoschiFilho:2002vd, BoschiFilho:2005yh,  Capossoli:2013kb, Rodrigues:2016cdb}. Note that the temperature of the black hole is related to the horizon position $z_H$, subjected to $z_H < z_{max}$.  

\begin{figure}[!h]
	\centering
	\includegraphics[scale = 0.6]{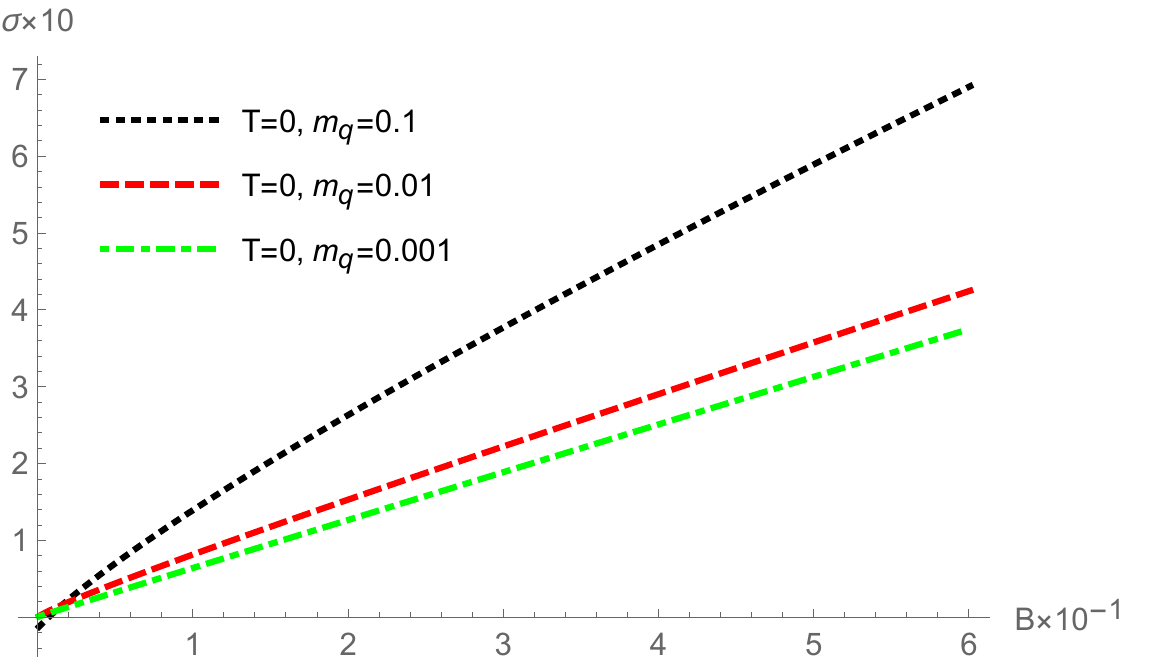}
	\caption{The chiral condensate $\sigma$ in units of $\sqrt{\sigma_s}$ versus the magnetic field $B$ in units of the string tension squared ${\sigma_s}$ in the confining regime at $ T = 0 $. 
		% Esta a figura Hardwall-figure4 (sigma-B at T=0.005) do Danning.
}
	\label{fig:t0-hardwall}
\end{figure}

\begin{figure}[!h]
	\centering
	\includegraphics[scale=0.6]{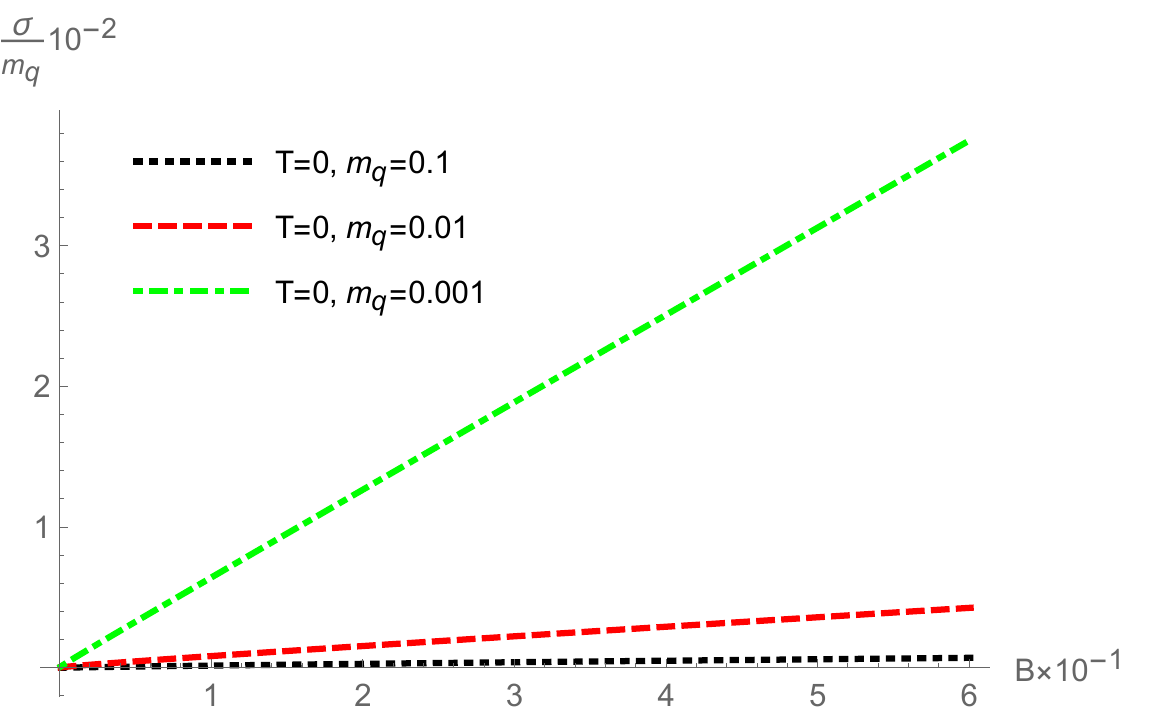}
	\caption{The chiral condensate $\sigma$ normalized by the fermion mass $m_q$  in units of $\sqrt{\sigma_s}$ versus the magnetic field $B$ in units of the string tension squared ${\sigma_s}$ in the confining regime at $ T = 0 $.}
	\label{fig:t0-rescaledhardwall}
\end{figure}

%\newpage
In Figure \ref{fig:t0-hardwall} we show the behavior of the chiral condensate as a function of the magnetic field at $ T = 0 $ for three fermion masses $\frac{m_q}{\sqrt{\sigma_s}}=0.001,\, 0.01, \, 0.1 $, where $\sqrt{\sigma_s}$ is the string tension. One can clearly see that in the confining regime the chiral condensate in the hardwall model is enhanced by the magnetic field yielding, as a result, MC. In Figure \ref{fig:t0-rescaledhardwall} we just plot the chiral condensate $\sigma$ normalized by the fermion mass $m_q$ versus the magnetic field $B$.

\begin{figure}[!h]
	\centering
	\includegraphics[scale=0.8]{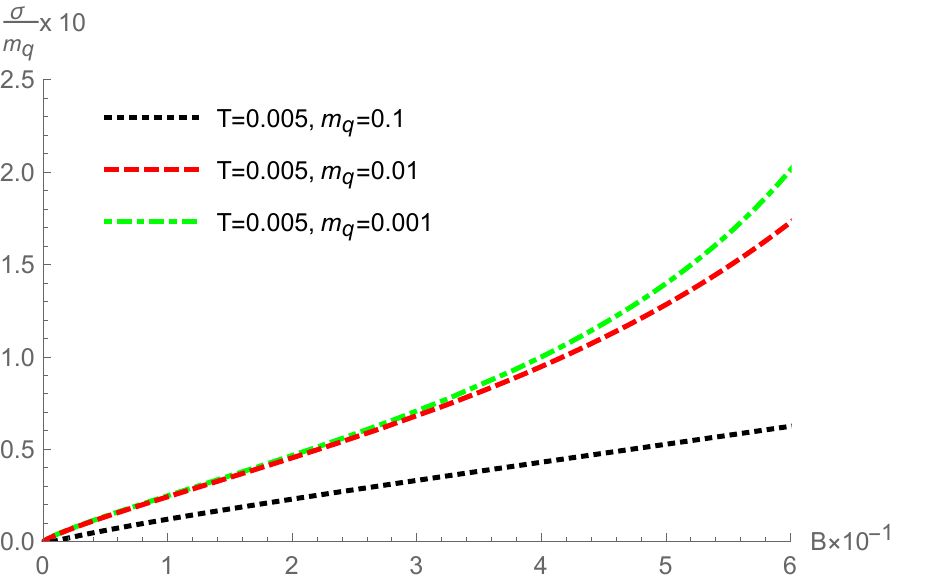}
	\caption{The chiral condensate $\sigma$ normalized by the fermion mass $m_q$  in units of $\sqrt{\sigma_s}$ versus the magnetic field $B$ in units of the string tension squared ${\sigma_s}$. This plot shows MC for low temperature $\frac{T}{\sqrt{\sigma_s}} = 0.005$ for three values of the fermion mass in the hardwall model. 
    % Esta a figura Hardwall-figure4 (sigma-B at T=0.005) do Danning.
    }
	\label{fig:1}
\end{figure}

In Figure \ref{fig:1}, at nonzero temperature but still in the confining phase, one continues to observe MC for low temperature $(\frac{T}{\sqrt{\sigma_s}} = 0.005)$ for the same three fermion masses $\frac{m_q}{\sqrt{\sigma_s}}=0.001,\, 0.01, \, 0.1 $. Also, in this picture one can note that for very weak magnetic fields the chiral condensate is almost independent of the fermion masses implying that, in the chiral limit ($m_q \rightarrow 0$), the chiral symmetry is spontaneously broken due to the presence of the magnetic field. As a last comment, for weak fields one can see an approximate linear behaviour for the chiral condensate consistent with the perturbative result found in \cite{Gusynin:1994re}, as can be seen in Eq. \eqref{cond}.  

%\newpage 

\begin{figure}[!h]
	\centering
	\includegraphics[scale=0.8]{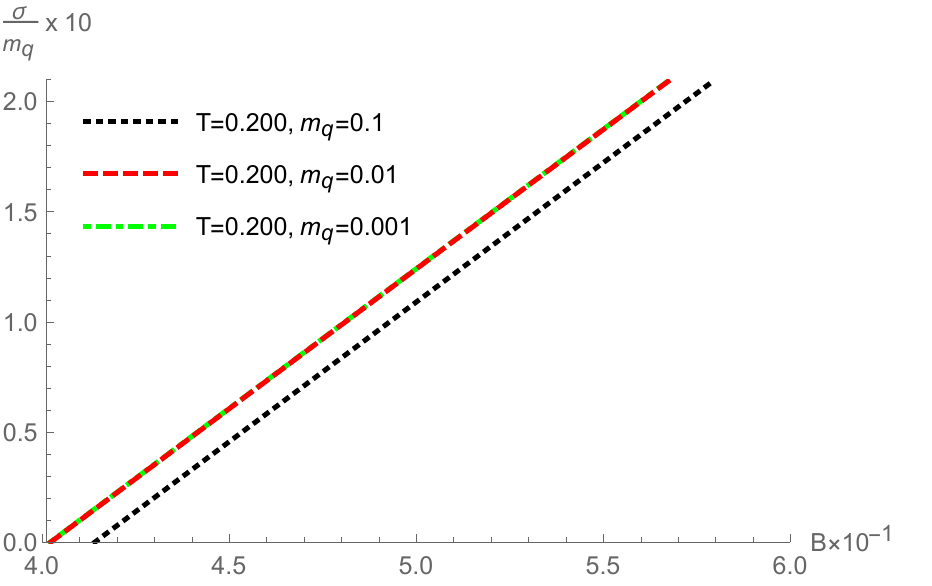}
	\caption{The chiral condensate $\sigma$ normalized by the fermion mass $m_q$  in units of $\sqrt{\sigma_s}$ versus the magnetic field $B$ in units of ${\sigma_s}$. This plot shows MC for the temperature $\frac{T}{\sqrt{\sigma_s}} = 0.200 $ for three values of the fermion mass in the hardwall model. 
    %Esta a figura Hardwall-figure3(sigma-B at T=0.2) do Danning.
    }
	\label{fig:2}
\end{figure}

In Figure \ref{fig:2} we can see that the increasing of the temperature still provides MC in the hardwall model. In this case the temperature is now 40 times greater $(\frac{T}{\sqrt{\sigma_s}} = 0.200)$ than in Figure \ref{fig:1}, and we still use the same values for the three fermion masses.  

%\newpage

\begin{figure}[!h]
	\centering
	\includegraphics[scale=0.6]{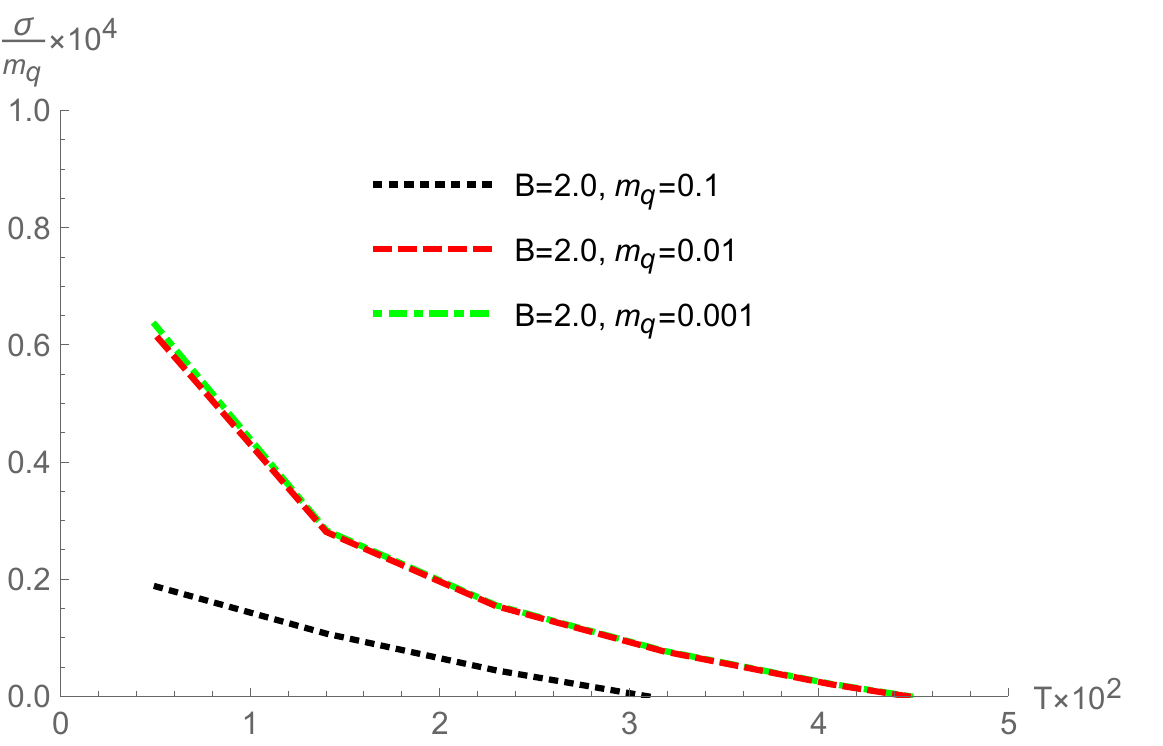}
	\caption{The chiral condensate $\sigma$ normalized by the fermion mass $m_q$ versus the temperature $T$, both in units of $\sqrt{\sigma_s}$. This plot also shows the restoration of the chiral symmetry  ($\sigma(T)=0$) for the hardwall model for a fixed and weak external magnetic field $\frac{B}{\sigma_s} = 2.0$ and different values of the fermion mass $m_q$. % and temperatures in the interval $[0.030 \, , \, 0.045]\, \sqrt{\sigma_s}$.
    %Esta a figura Hardwall-figure6(sigma-T at B=2.0) do Danning.
    }
	\label{fig:3}
\end{figure}

\begin{figure}[!h]
	\centering
	\includegraphics[scale=0.6]{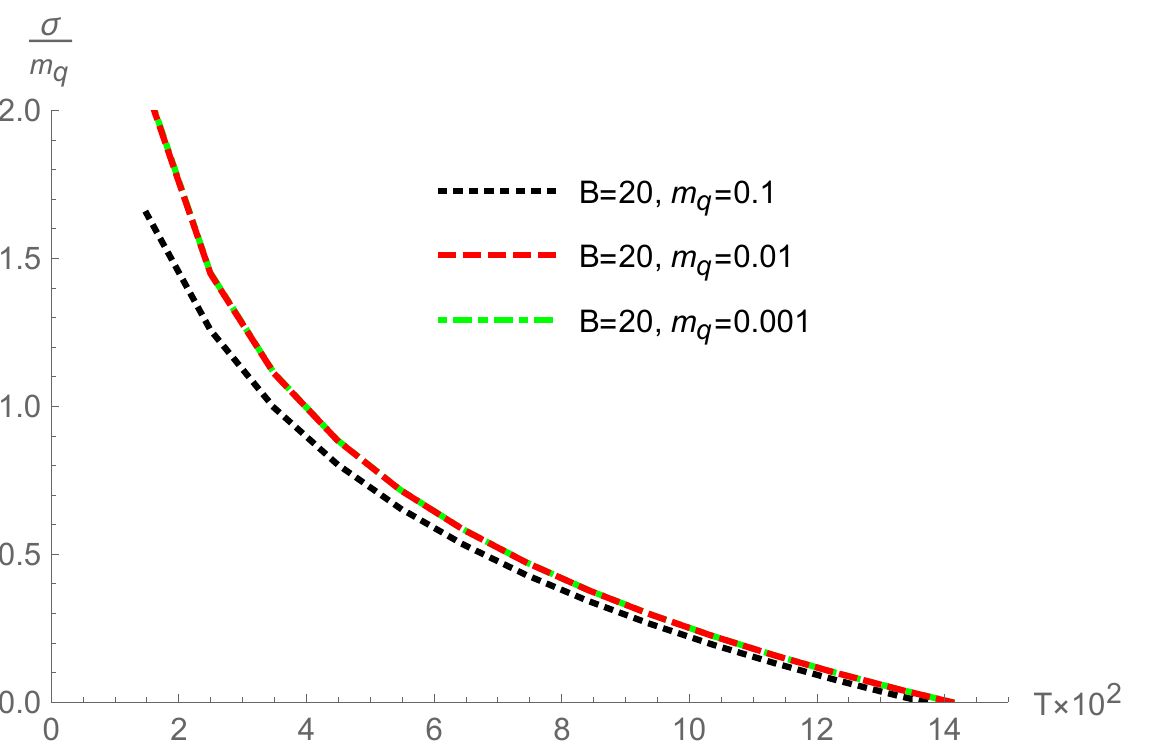}
	\caption{The chiral condensate $\sigma$ normalized by the fermion mass $m_q$ versus the temperature $T$, both in units of $\sqrt{\sigma_s}$. This plot also shows the restoration of the chiral symmetry ($\sigma(T)=0$) for the hardwall model for a fixed and strong external magnetic field $\frac{B}{\sigma_s} = 20$ and different values of the fermion mass $m_q$. %and temperature of the order of $\frac{T}{\sqrt{\sigma_s}}\sim 0.14 $. 
    %Esta  figura nao foi salva como pdf, mas esta no nb Hardwall-figure5,6.
    }
	\label{fig:4}
\end{figure}

In Figures \ref{fig:3} and \ref{fig:4} we present the dependence of the chiral condensate $\sigma$ with respect to the temperature $T$. These figures show the restoration of the chiral symmetry ($\sigma \to 0$) for different values of the external magnetic field and of the fermion mass $m_q$. 

%\newpage

One can also note that by comparison between Figures \ref{fig:3} and \ref{fig:4} the MC behavior since the values of the chiral condensate increases when the magnetic field is also increased as shown explicitly in Figures \ref{fig:1} and \ref{fig:2}. 

\begin{figure}[h]
	\centering
	\includegraphics[scale=0.6]{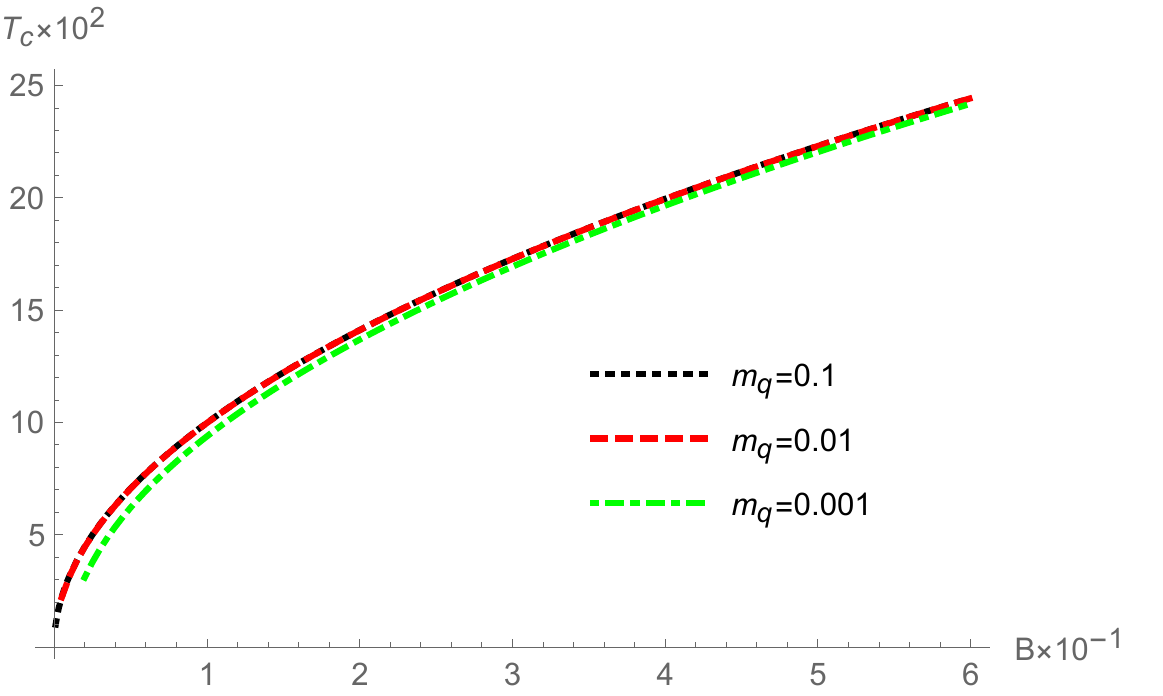}
	\caption{The critical temperature, $T_c$, in units of $\sqrt{\sigma_s}$, as a function of the magnetic field, $B$, in units of $ \sigma_s $ for the hardwall model for different values of the fermion mass.}
	\label{fig:tc-hardwall}
\end{figure}

We also show in Figure \ref{fig:tc-hardwall} the behaviour of the critical temperature $(T_c)$ against the magnetic field ($B$) for the hardwall model. Here, we use the definition of $T_c$ as the temperature where $\sigma(T)=0$, like in ref. \cite{Dudal:2015wfn}. Note that $T_c$ increases as we increase the magnetic field as a consequence of the MC observed, for instance, in Fig. \ref{fig:1}. 

\clearpage

\section{Results for the Softwall Model}

In this section we do the entire analysis of the previous section this time for the softwall model  \cite{Karch:2006pv, Colangelo:2007pt,FolcoCapossoli:2016ejd}, characterized by a dilatonlike field $\Phi(z)$.  We assume the dilaton profile  to be of the form $\Phi(z) = kz^2$, where the constant $k$ was set as $\frac{k}{(\sqrt{\sigma_s})^2} = -1$. Further, one can note that the soft wall linear spectrum persists in 2+1 dimensions, as has been shown in the Appendix of ref. \cite{Rodrigues:2017iqi}. There, one uses the dilaton constant k in units of the string tension. In this model we are also assuming that there is no backreaction due to the dilatonlike field so that the equations of motions for this model are still given by \eqref{FieldEquations}, whose solution corresponds to  \eqref{fBT}. In the following we present our results for this model.

\begin{figure}[!h]
	\centering
	\includegraphics[scale=0.4]{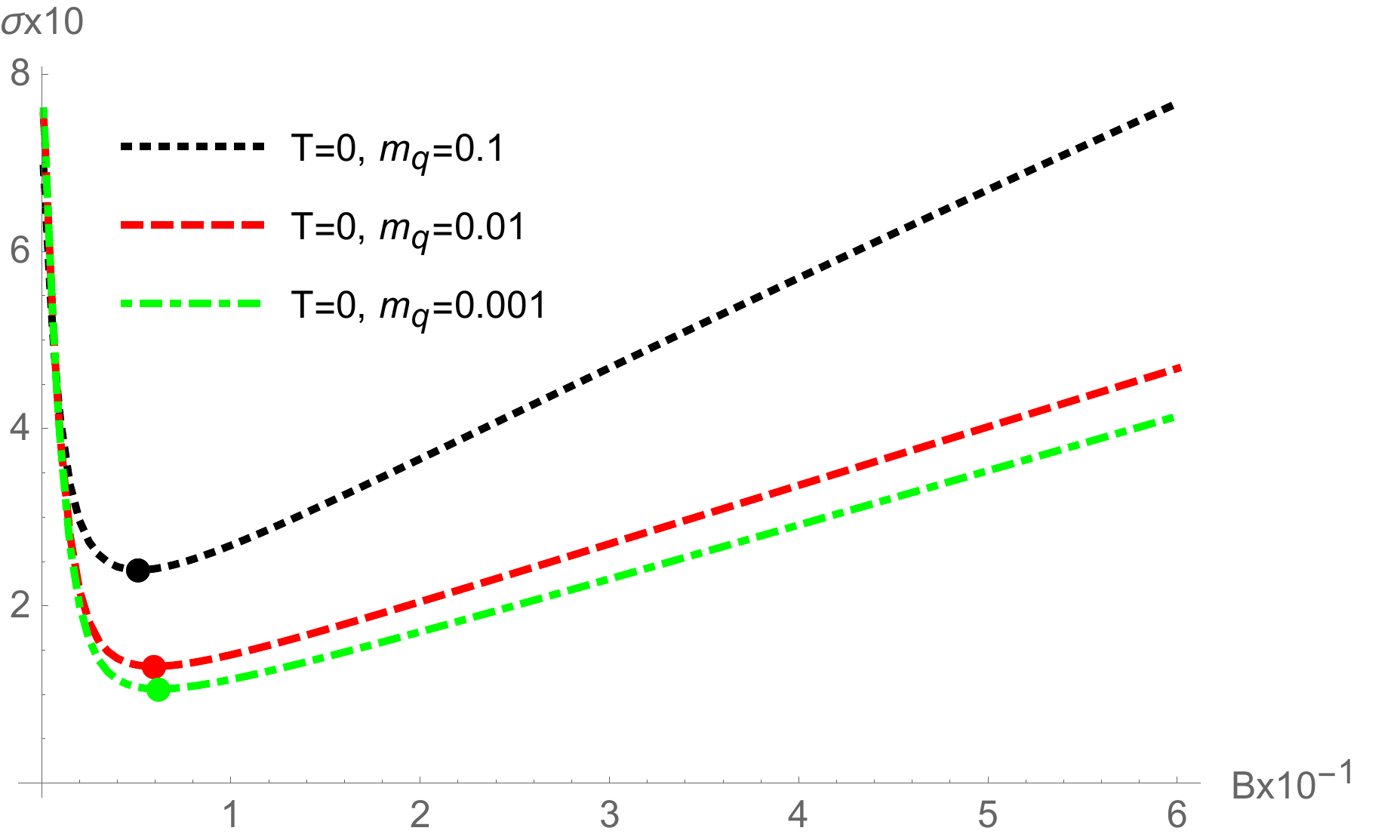}
	\caption{The chiral condensate $\sigma$ versus the magnetic field $B$, both in units of ${\sigma_s}$. This plot shows IMC for $B<B_c$ and MC for $B>B_c$ for the softwall model at zero temperature ($T = 0$) and three fermion masses. The dots in the lines represent the pseudocritical magnetic field $B_c$, which characterizes a crossover between the IMC and MC phases. In particular, one can read $\frac{B_c}{\sigma_s} = 6.17$ for $\frac{m_q}{\sqrt{\sigma_s}} = 0.001$;  $\frac{B_c}{\sigma_s} = 5.94$ for $\frac{m_q}{\sqrt{\sigma_s}} = 0.01$; and $\frac{B_c}{\sigma_s} = 5.12$ for $\frac{m_q}{\sqrt{\sigma_s}} = 0.1 $.}
	\label{fig:t0swdisk}
\end{figure}

We begin by presenting the behavior of the chiral condensate $\sigma$ as a function of the magnetic field $ B $ in the confining regime ($ T=0 $) for three fermion masses $m_q$. This result is shown in Figure \ref{fig:t0swdisk}. Unlike the results for the hardwall model in the previous section, one observes that the chiral condensate in the softwall model presents two different behaviors depending on the value of the magnetic field. These behaviors are separated by a pseudo critical magnetic field, $ B_c $, such that for $B<B_c $ we have a IMC phase and for $B>B_c $ we have a MC phase. 

\begin{figure}[!h]
	\centering
	\includegraphics[scale=0.4]{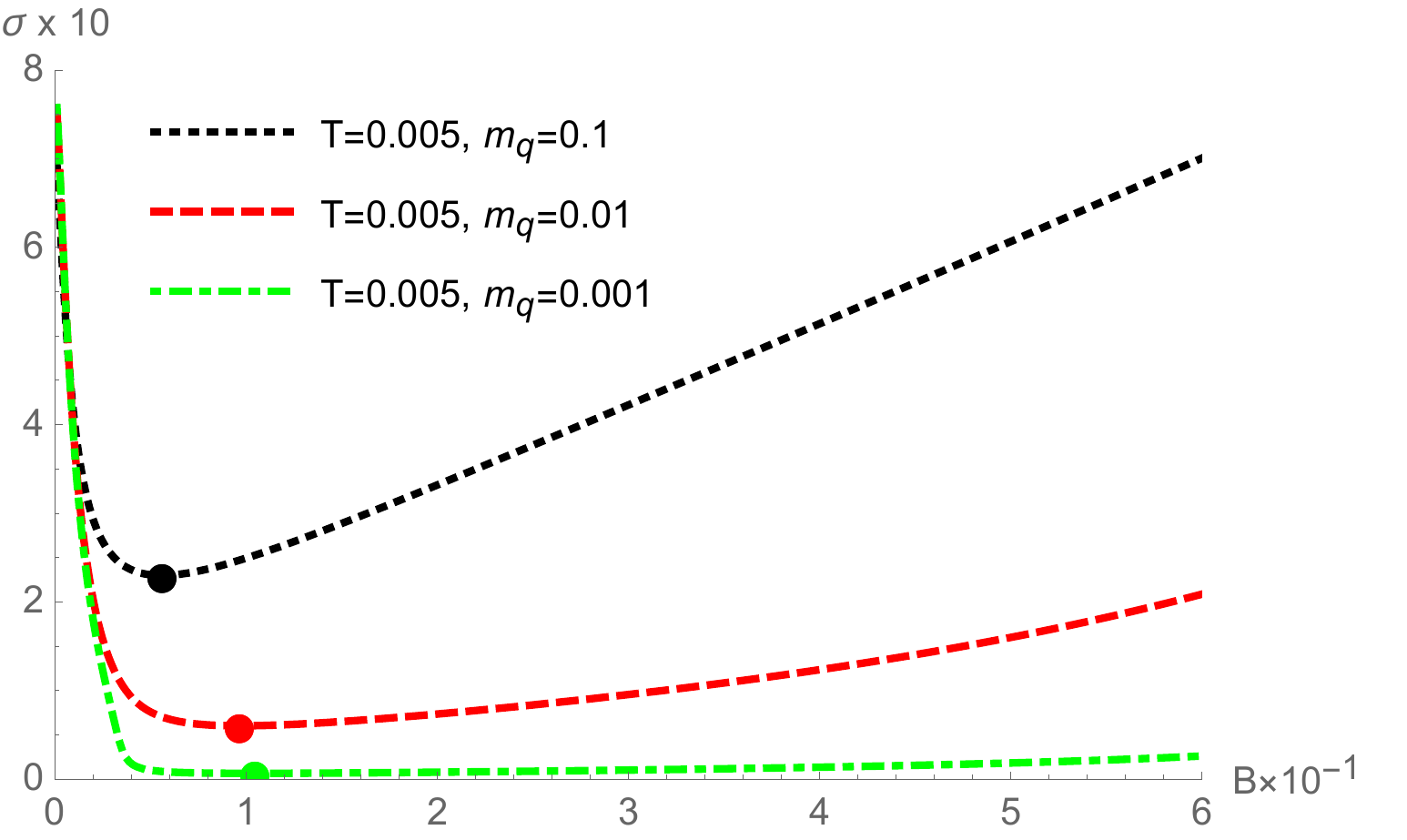}
	\caption{The chiral condensate $\sigma$ versus the magnetic field $B$, both in units of ${\sigma_s}$. This plot shows IMC for $B<B_c$ and MC for $B>B_c$ for the softwall model with fixed temperature $\frac{T}{\sqrt{\sigma_s}} = 0.005$ and three fermion masses. The dots in the lines represent the pseudocritical magnetic field $B_c$, which characterizes a crossover between the IMC and MC phases. In particular, one can read $\frac{B_c}{\sigma_s} = 10.44$ for $\frac{m_q}{\sqrt{\sigma_s}} = 0.001$;  $\frac{B_c}{\sigma_s} = 9.67$ for $\frac{m_q}{\sqrt{\sigma_s}} = 0.01$; and $\frac{B_c}{\sigma_s} = 5.59$ for $\frac{m_q}{\sqrt{\sigma_s}} = 0.1 $.}
	\label{fig:6}
\end{figure}

\begin{figure}[!h]
	\centering
	\includegraphics[scale=0.65]{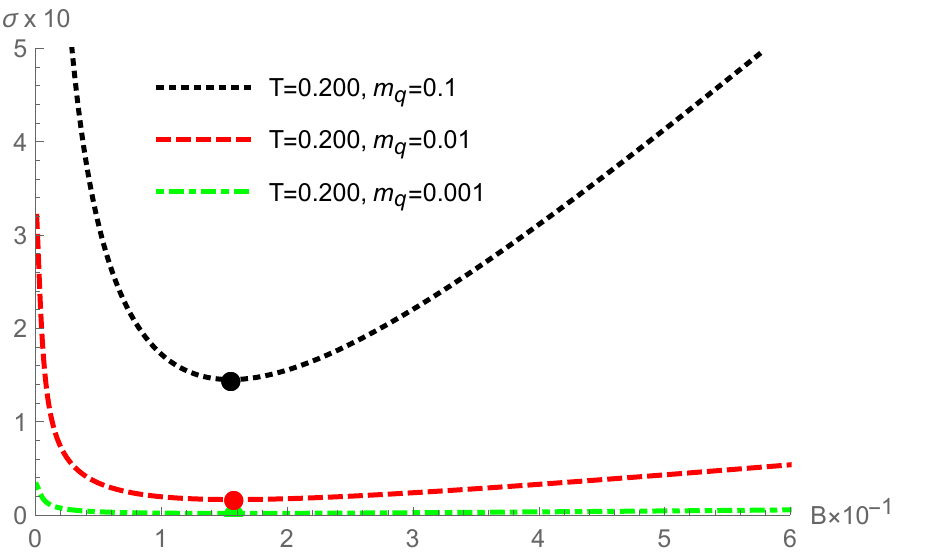}
	\caption{The chiral condensate $\sigma$ versus the magnetic field $B$, both in units of ${\sigma_s}$. This plot shows IMC for $B<B_c$ and MC for $B>B_c$ for the softwall model with fixed temperature $\frac{T}{\sqrt{\sigma_s}} = 0.200 $ and three fermion masses. The dots in the lines represent the pseudocritical magnetic field $B_c $, which characterizes a crossover between the IMC and MC phases. In particular, one can read $\frac{B_c}{\sigma_s} = 15.78$ for $\frac{m_q}{\sqrt{\sigma_s}} = 0.001$;  $\frac{B_c}{\sigma_s} = 15.77$ for $\frac{m_q}{\sqrt{\sigma_s}} = 0.01$; and $\frac{B_c}{\sigma_s} = 15.52$ for $\frac{m_q}{\sqrt{\sigma_s}} = 0.1 $.}
	\label{fig:7}
\end{figure}

Figures \ref{fig:6} and \ref{fig:7} show IMC and MC for the softwall model at finite temperature for both low and high temperatures and three fermion masses $\frac{m_q}{\sqrt{\sigma_s}} = 0.001, \, 0.01,\, 0.1$. 
In Figure \ref{fig:6} the temperature is $\frac{T}{\sqrt{\sigma_s}} = 0.005$, while in Figure \ref{fig:7} the temperature is $\frac{T}{\sqrt{\sigma_s}} = 0.200$. One should note that these two figures present a crossover between the two phases IMC and MC around a pseudocritical magnetic field $B_c$ which depends on the temperature. Note that the comparison between Figures \ref{fig:6} and \ref{fig:7} also shows implicitly a chiral symmetry restoration since $\sigma$ decreases for an increasing temperature, for $B<B_c$. 

%

%\newpage 

%\newpage

\begin{figure}[!h]
	\centering
	\includegraphics[scale=0.5]{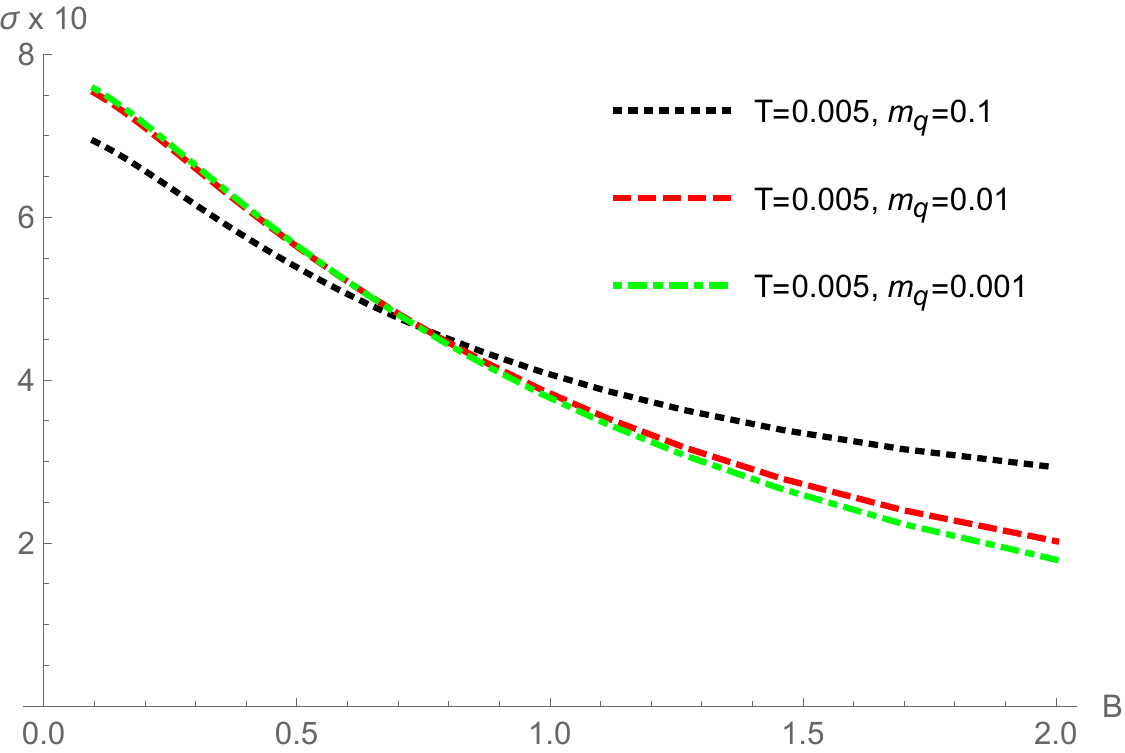}
	\caption{The chiral condensate $\sigma$ versus the magnetic field $B$, both in units of ${\sigma_s}$ for the softwall model. This plot shows IMC for weak magnetic fields and low temperature. This plot also shows spontaneous breaking of the chiral symmetry. The independence of the fermion mass for the chiral condensate is more evident for $\frac{m_q}{\sqrt{\sigma_s}} \leq 0.01$.   
 %Esta figura foi contruida do arquivo 2+1D chiral-data-psoft-nsoft-hard-2. 
    }
	\label{fig:11}
\end{figure}

\begin{figure}[!h]
	\centering
	\includegraphics[scale=0.5]{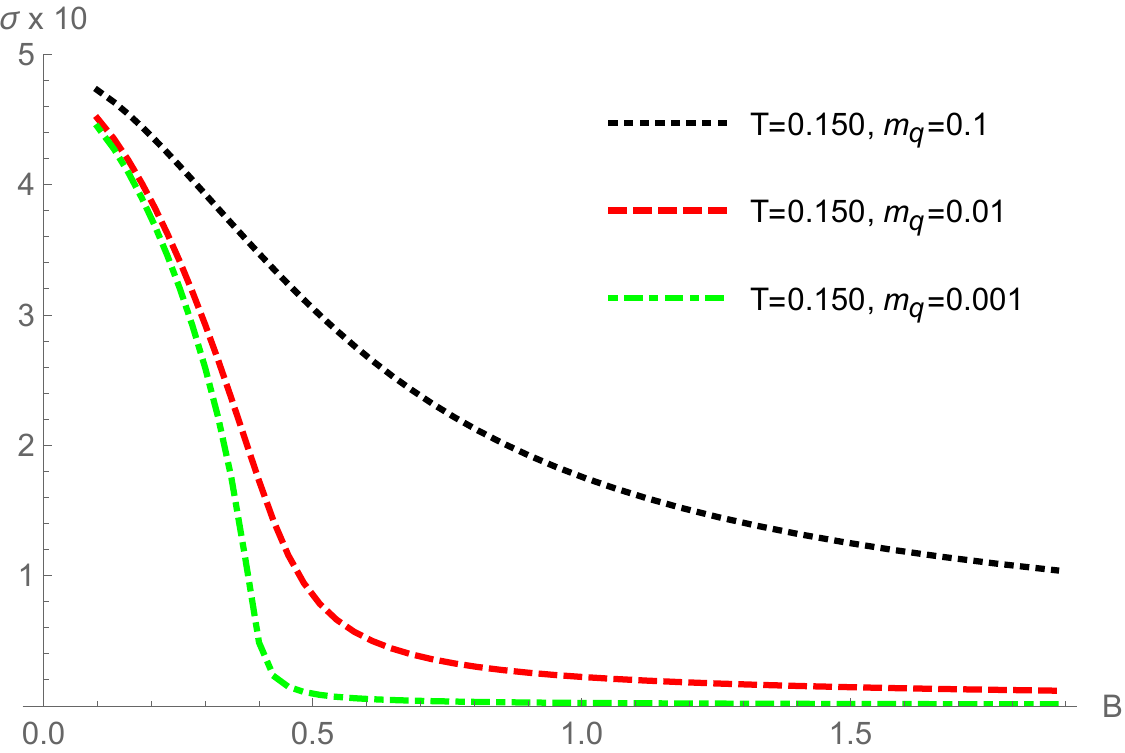}
	\caption{The chiral condensate $\sigma$ versus the magnetic field $B$, both in units of ${\sigma_s}$ for the softwall model with temperature $\frac {T}{\sqrt{\sigma_s}}=0.150$. This plot shows IMC for weak magnetic fields and high temperature and spontaneous breaking of the chiral symmetry. The approximate independence of the fermion mass for the chiral condensate occurs for $\frac{m_q}{\sqrt{\sigma_s}} \leq 0.01$. 
 %Esta figura foi contruida do arquivo 2+1D chiral-data-psoft-nsoft-hard-2. 
    } 
	\label{fig:12}
\end{figure}

\begin{figure}[!h]
	\centering
	\includegraphics[scale=0.5]{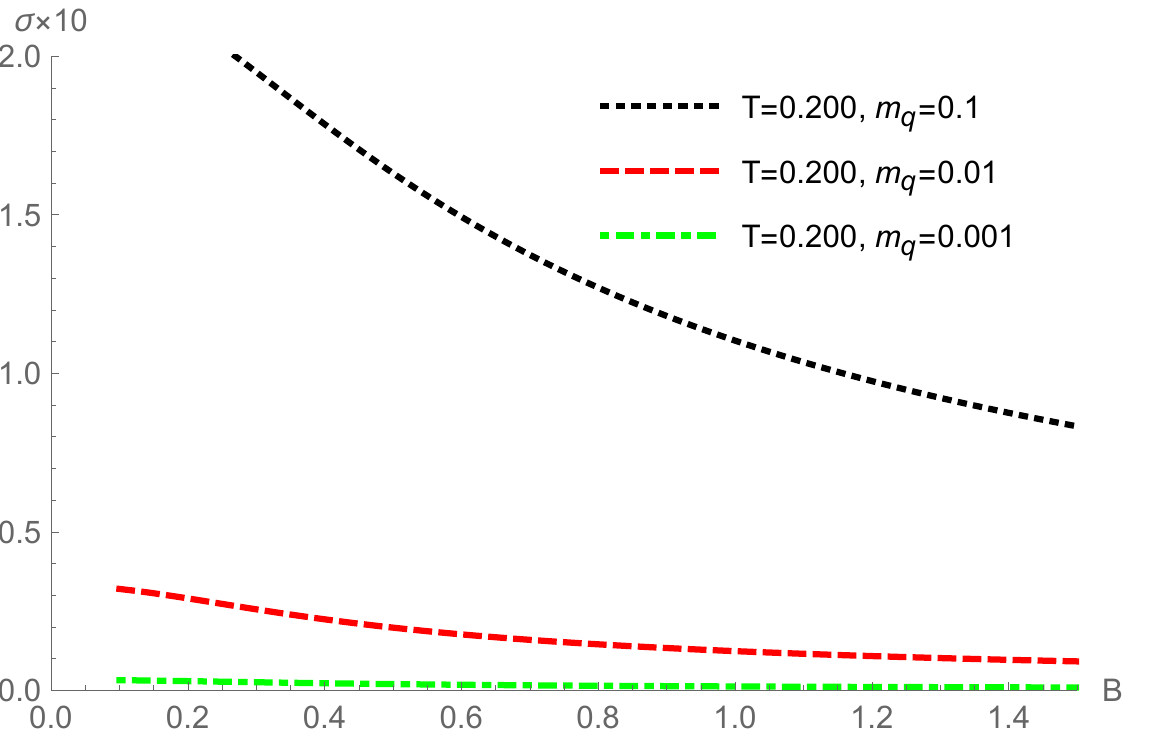}
	\caption{The chiral condensate $\sigma$ versus the magnetic field $B$, both in units of ${\sigma_s}$ for the softwall model with temperature $\frac {T}{\sqrt{\sigma_s}}=0.200$. This plot shows IMC for weak magnetic fields and high temperature and spontaneous breaking of the chiral symmetry. The approximate independence of the fermion mass for the chiral condensate is more evident for $\frac{m_q}{\sqrt{\sigma_s}} \leq 0.01$.
 %Esta figura foi contruida do arquivo 2+1D chiral-data-psoft-nsoft-hard-2. 
    } 
	\label{fig:12_new}
\end{figure}

The Figures \ref{fig:11}, \ref{fig:12} and \ref{fig:12_new} represent closer looks for the chiral condensate behavior in the region of weak magnetic fields.
In Figure \ref{fig:11} for very low temperature, one can note that the chiral condensate's behavior is approximately the same for the three values of the fermion mass, indicating a spontaneous chiral symmetry breaking. On the other hand, as one increases the temperature as shown in Figures \ref{fig:12} and \ref{fig:12_new}, its effect on the chiral condensate is more visible for the lowest fermion masses $(\frac{m_q}{\sqrt{\sigma_s}} \le 0.01)$, close to the chiral limit. In this case we still have an approximate chiral symmetry breaking. Comparing Figures \ref{fig:12} and \ref{fig:12_new}, one can see that a small increase of the temperature from $\frac{T}{\sqrt{\sigma_s}} = 0.150 $ to $\frac{T}{\sqrt{\sigma_s}} = 0.200 $ affects the lighter fermions masses more. 


%\newpage

\begin{figure}[!h]
	\centering
	\includegraphics[scale=0.4]{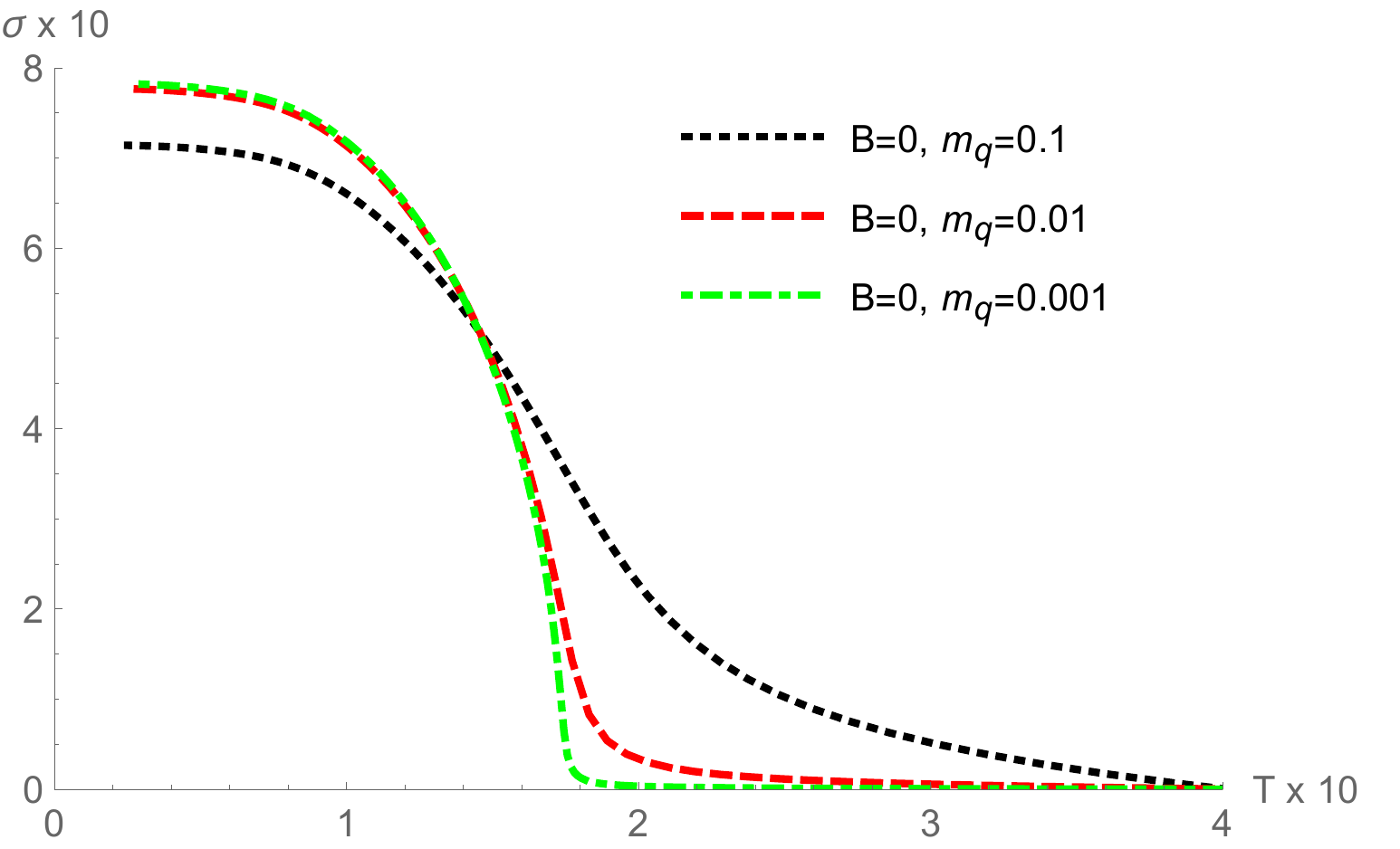}
	\caption{The chiral condensate $\sigma$, in units of ${\sigma_s}$, versus the temperature $T$, in units of $\sqrt{\sigma_s}$ for the softwall model. This plot shows the chiral symmetry restoration ($\sigma(T)=0$) for different values of the fermion mass without external magnetic field. This picture also shows an approximate independence of the chiral condensate with the fermion mass, in particular for $\frac{m_q}{\sqrt{\sigma_s}} \leq 0.01$. %The temperature of the restoration corresponds approximately to the interval $[0.2 \, ,\, 0.4 ]$, in units of the string tension.   
    %Esta a figura Negative-figure new 2(sigma-T at B=0) do Danning. 
    }
	\label{fig:8}
\end{figure}

\begin{figure}[!h]
	\centering
	\includegraphics[scale=0.4]{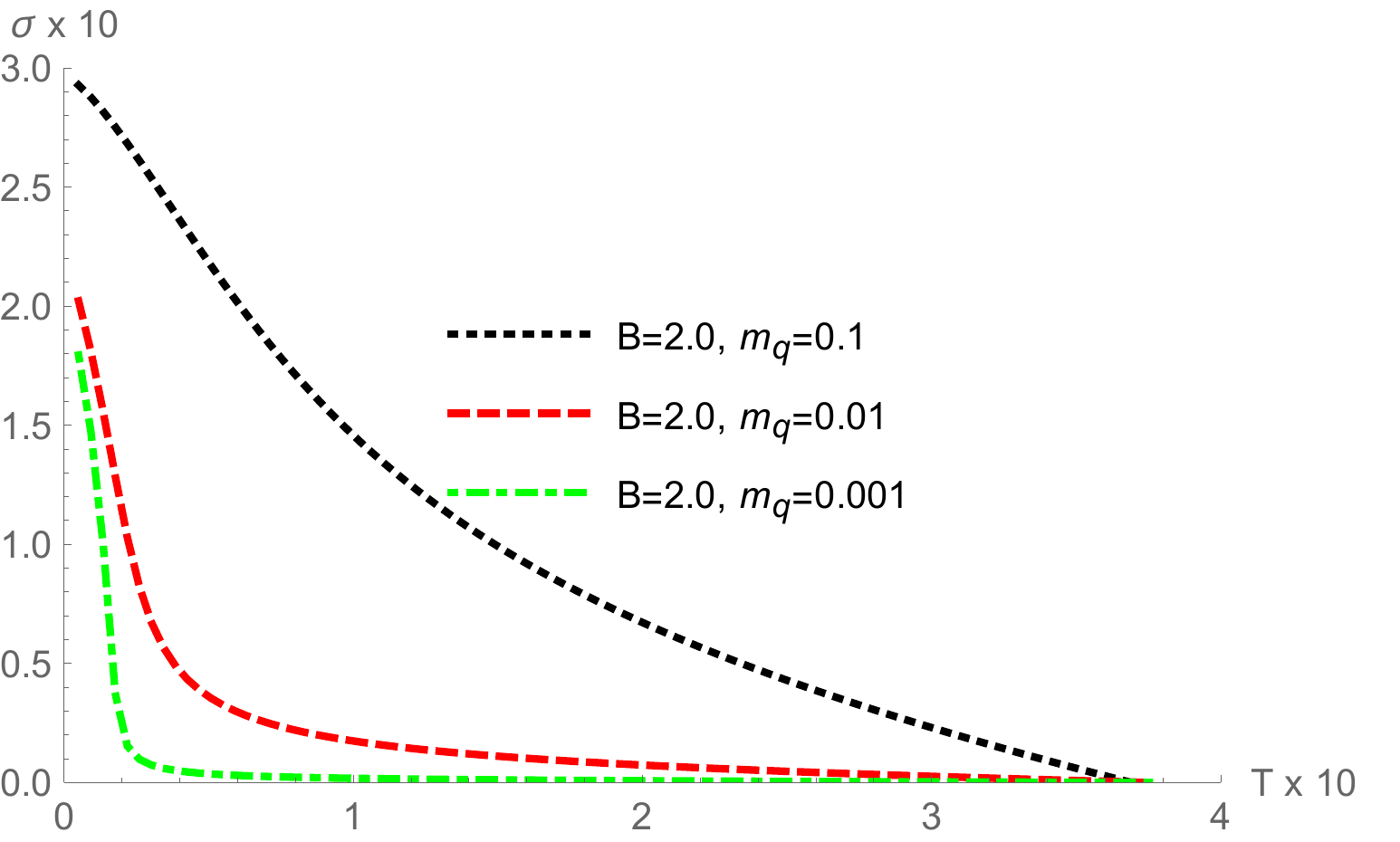}
	\caption{The chiral condensate $\sigma$, in units of ${\sigma_s}$, versus the temperature $T$, in units of $\sqrt{\sigma_s}$ for the softwall model. This plot shows the chiral symmetry restoration ($\sigma(T)=0$) for different values of the fermion mass with external magnetic field $\frac{B}{\sigma_s} = 2.0$. This picture also shows an approximate independence of the chiral condensate with the fermion mass, in particular for $\frac{m_q}{\sqrt{\sigma_s}} \leq 0.01$. %The temperature of the restoration corresponds  approximately to the interval $[0.02 \, ,\, 0.4 ]$, in units of the string tension. 
  %  Esta a figura Negative-figure6(sigma-T at B=2.0) do Danning.
    }
	\label{fig:9}
\end{figure}

\begin{figure}[!h]
	\centering
	\includegraphics[scale=0.4]{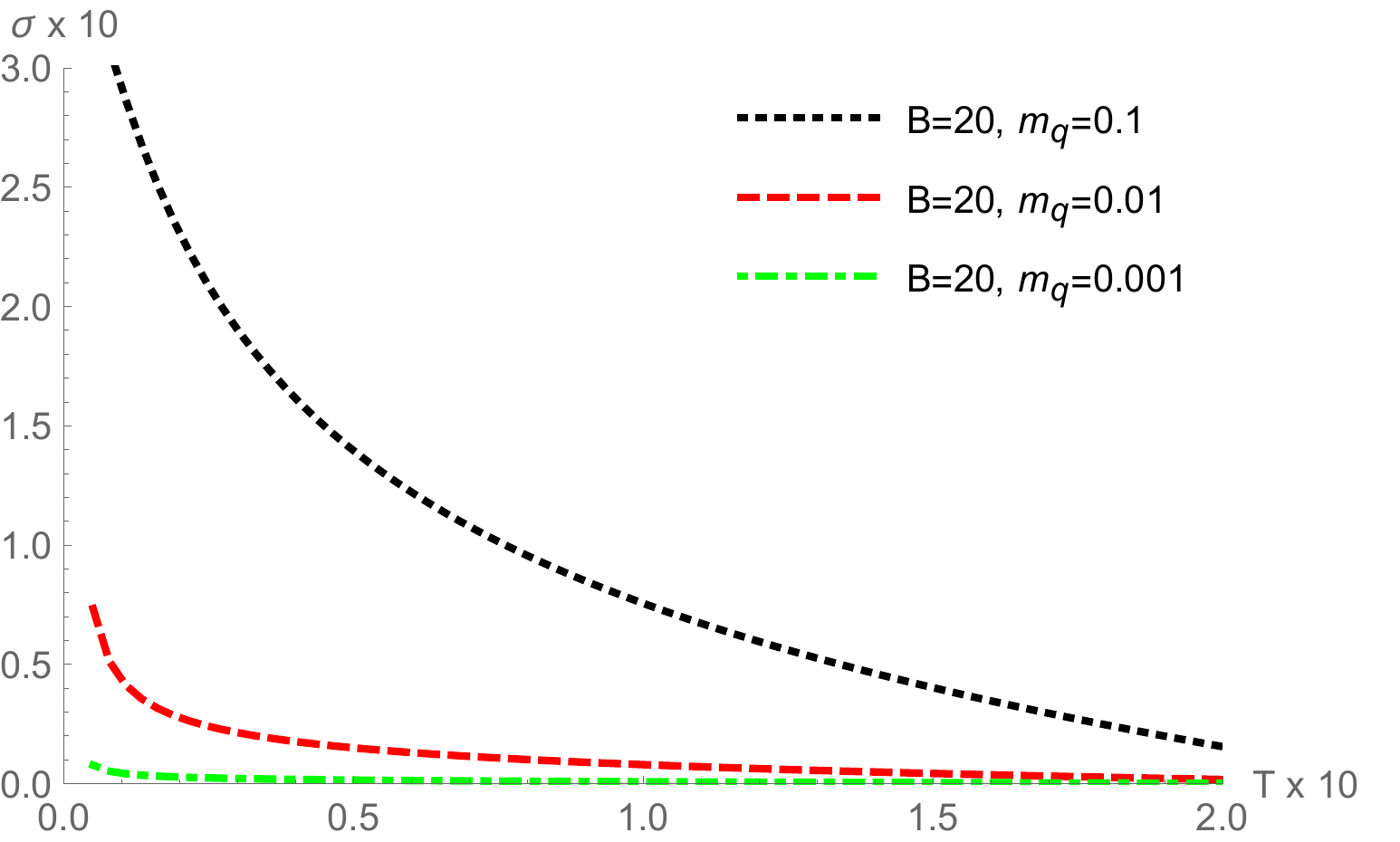}
	\caption{The chiral condensate $\sigma$, in units of ${\sigma_s}$, versus the temperature $T$, in units of $\sqrt{\sigma_s}$ for the softwall model. This plot shows the chiral symmetry restoration ($\sigma(T)=0$) for different values of the fermion mass with external magnetic field $\frac{B}{\sigma_s} = 20$. This picture also shows an approximate independence of the chiral condensate with the fermion mass, in particular for $\frac{m_q}{\sqrt{\sigma_s}} \leq 0.01$. %The temperature of the restoration corresponds  approximately to the interval $[0.01 \, ,\, 0.2 ]$, in units of the string tension.  
    %Esta a figura Negative-figure new 1(sigma-T at B=20.0) do Danning.
    }
	\label{fig:10}
\end{figure}

Figures \ref{fig:8}, \ref{fig:9} and \ref{fig:10} show the behavior of the chiral condensate with respect to the temperature for different values of the fermion mass and different values of the magnetic field. All three figures show the restoration of the chiral symmetry as one increases the temperature, so that the chiral condensate goes to zero. In particular, in Fig. \ref{fig:8}, one notes that even at zero external magnetic field this model provides a nonzero value for the chiral condensate at low temperatures. This is in contrast with the perturbative result, Eq. \eqref{cond}, so one could  associate this behavior with a nonperturbative effect. 
However previous studies concerning the chiral symmetry breaking in $ (2+1) $ dimensions without magnetic field ($ B=0 $) were done in the context of QED$ _3 $, within the large $ N $ approximation \cite{Appelquist:1985vf,Appelquist:1986fd}. There it was shown that it is energetically preferable for this theory to dynamically generate a mass for fermions. In a subsequent analysis by the same authors it was shown, in the lowest order in the large $ N $ approximation, that the dynamical chiral symmetry breaking happens only when the number of fermions $ N $ is less than a critical number $N_{c} = 32/\pi^2 $ \cite{Appelquist:1988sr}. After that, in \cite{Nash:1989xx}, it was shown that higher order corrections do not modify the nature of the chiral symmetry breaking. 

\begin{figure}
	\centering
	\includegraphics[scale = 0.6]{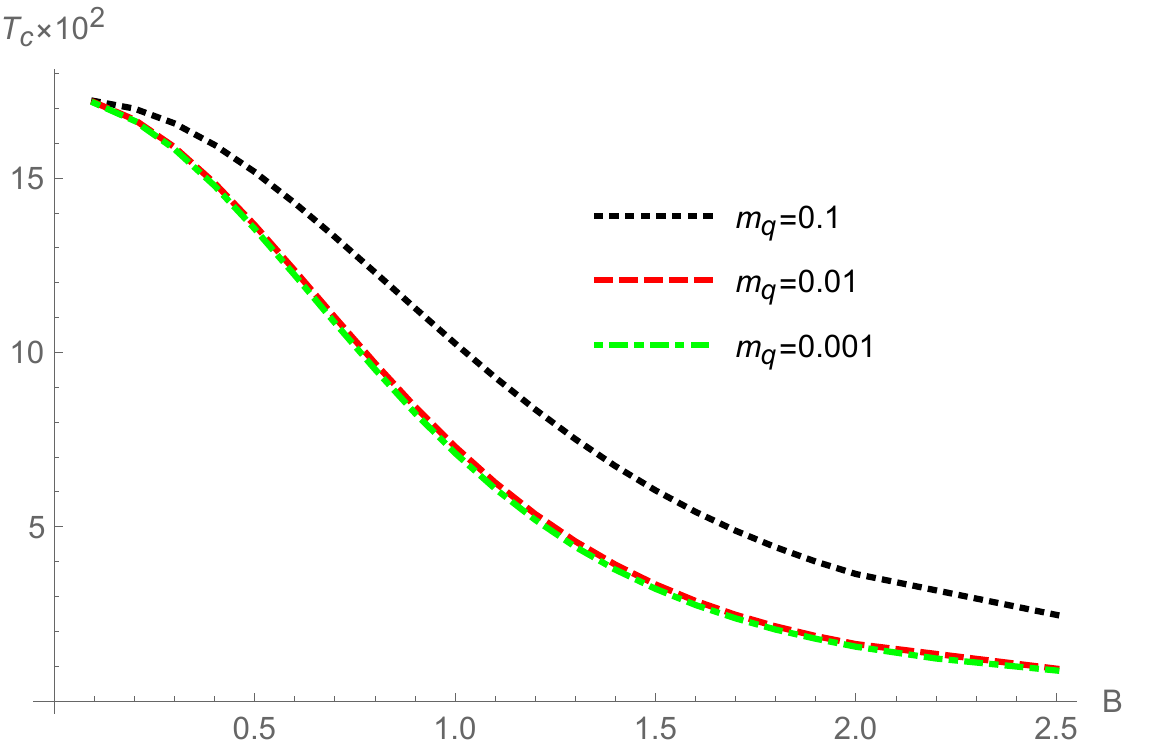}
	\caption{Critical temperature, $T_c$, in units of $\sqrt{\sigma_s}$, as a function of the magnetic field, $B$, in units of $ \sigma_s $, for the softwall model in the IMC phase (low $B$ regime), for different values of the fermion mass.}
	\label{fig:tc-negative-softwall}
\end{figure}

\begin{figure}
	\centering
	\includegraphics[scale = 0.6]{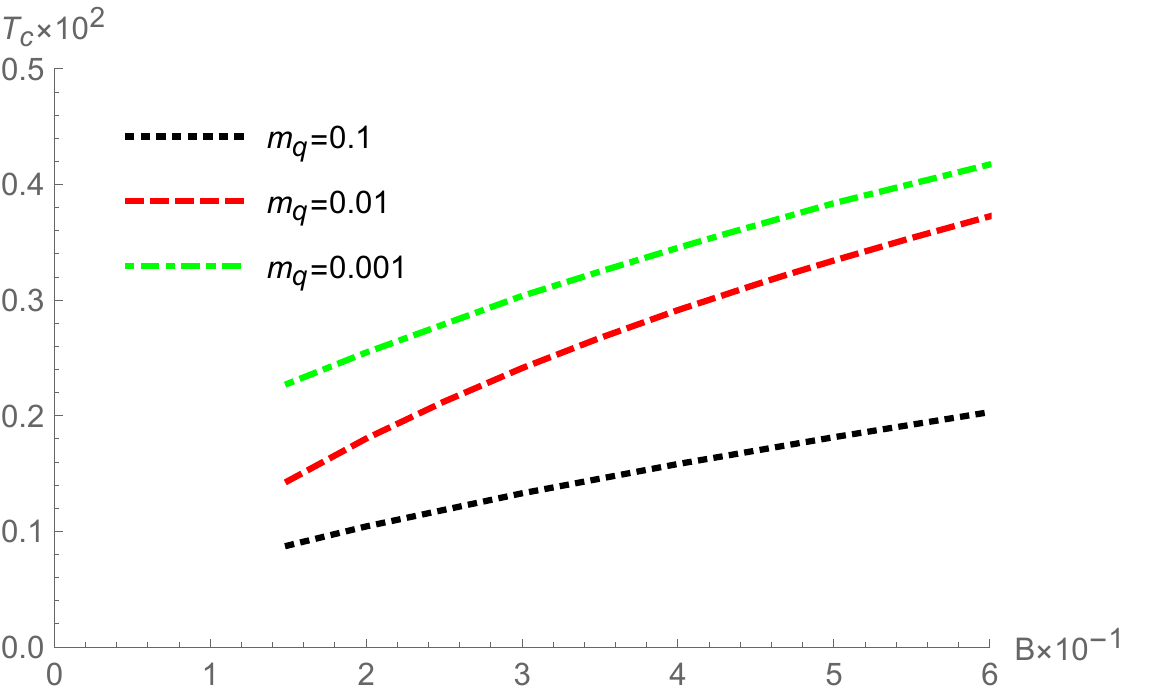}
	\caption{Critical temperature, $T_c$, in units of $\sqrt{\sigma_s}$, as a function of the magnetic field, $B$, in units of $ \sigma_s $, for the softwall model in the MC phase (high $B$ regime), for different values of the fermion mass.}
	\label{fig:tc-b1560-negative-softwall}
\end{figure}

We also plot in Figures \ref{fig:tc-negative-softwall} and \ref{fig:tc-b1560-negative-softwall} the behaviour of the critical temperature $(T_c)$ against the magnetic field ($B$). In this case for the softwall model $T_c$ is defined such that $\sigma(T)$ decreases fastest (maximum of $-d\sigma(T)/dT$), as in ref.  \cite{Chelabi:2015gpc}. Note that for weak fields, Fig. \ref{fig:tc-negative-softwall}, $T_c$ decreases as we increase the magnetic field, as a consequence of the IMC, as for instance, in Figs. \ref{fig:6} and \ref{fig:7}. On the other hand, in Fig. \ref{fig:tc-b1560-negative-softwall}, for strong fields, $T_c$ increases as we increase the magnetic field as a consequence of MC, as for instance in Figs. \ref{fig:6} and \ref{fig:7}. 

%\newpage

%\newpage

\clearpage

\section{Discussion and Conclusion}

In this work we have described holographically the chiral phase transition in $(2+1)$ dimensions in the presence of an external magnetic field $B$ at finite temperature $T$. We have used two well-known holographic models: the hardwall and the softwall. In the first one we have observed just magnetic catalysis (MC), that is, the chiral condensate $\sigma$ is enhanced due to the presence of the magnetic field, and is zero in the absence of magnetic field, in qualitative agreement with the perturbative analysis in \cite{Gusynin:1994re,Shovkovy:2012zn,Miransky:2015ava}. Moreover, in the region of small magnetic field, the behavior of the chiral condensate as function of the magnetic field seems to be approximately linear with $B$. This can be seen in Figs. \ref{fig:t0-hardwall}, \ref{fig:t0-rescaledhardwall} and \ref{fig:1} for very low $T$. This is also in agreement with Eq. \eqref{cond}, derived from a perturbative quantum field theory approach for free fermions in a magnetic field in $(2+1)$ dimensions at $T=0$. Although these results for the hardwall model look very similar to the perturbative one, we considered in this work a nonperturbative approach, since we are in the framework of the AdS/CFT correspondence. Moreover, our results for the hardwall model agree with the one found in \cite{Dudal:2015wfn}, in the context of the holographic QCD in ($3+1$) dimensions. There the authors concluded that there is no IMC for the hardwall, but instead what they obtained was just MC.

Our results within the softwall model, on the other hand, reveal some new interesting features. First of all, we obtained chiral symmetry breaking even for zero magnetic field ($B=0$), as can be seen in Fig. \ref{fig:8}, completely in contrast with the perturbative result described by Eq. \eqref{cond}. This fact seems to be associated with the nonpertubative feature of our holographic approach. Also, we have concluded, for zero and nonzero $B$, that the breaking of the chiral symmetry is a spontaneous one, since the dependence on the chiral condensate $\sigma$ with the magnetic field $B$ for different fermion masses $m_q$ is approximately the same, as can be seen in the plots of $\sigma$ x $B$, represented by Figures \ref{fig:11}, \ref{fig:12} and \ref{fig:12_new}. Remarkably, the softwall model provides a crossover between the IMC/MC phases as shown in the plots of $\sigma$ x $B$, Figures \ref{fig:t0swdisk}, \ref{fig:6} and \ref{fig:7}, characterized by a pseudocritical magnetic field $B_c$ for each fermion mass for a given temperature. Therefore, in this sense, for $B > B_c$ we recover approximately the linear behaviour $\sigma \sim B$ as in the perturbative result given by \eqref{cond}. 
On the other hand, for $B<B_c$, we found IMC which is in qualitative agreement with those found in lattice QCD \cite{Bali:2011qj,Endrodi:2015oba} and the holographic approaches dealing with QCD in $(3+1)$ dimensions \cite{Li:2016gfn,Chelabi:2015cwn,Chelabi:2015gpc}. Note that all of these features that our holographic model presents in $(2+1)$ dimensions were not observed in the other holographic approaches for QCD because their results are perturbative in $B$. Therefore they could trust the numerical results only for small $B$, while ours is exact in the magnetic field, meaning that it is valid for any $B$. 

Now, concerning the thermal effects on the chiral condensate $\sigma$, as can be seen in the plots of $\sigma$ x $T$, Figures \ref{fig:3} and \ref{fig:4} for the hardwall and  Figures \ref{fig:8}, \ref{fig:9} and \ref{fig:10} for the softwall, 
we have observed that the restoration of the chiral symmetry ($\sigma=0$) happens at some critical temperature $T_c$ for each model, depending on the fermion mass and on the magnetic field (see also Figs. \ref{fig:tc-hardwall},         
\ref{fig:tc-negative-softwall} and \ref{fig:tc-b1560-negative-softwall}). This phenomenon  is consistent with lattice QCD results and with the holographic approaches for QCD in (3+1) dimensions, previously mentioned. 

Finally, in comparison with the perturbative counterpart at finite temperature $T$ in ($2+1$) dimensions \cite{Das:1995bn}, in which the authors have shown that the chiral condensate $\sigma$ is extremely unstable (it vanishes as soon as one introduces a heat bath, with or without a chemical potential), we have seen that in the models we have considered the chiral condensate persists up to some critical temperature, and that instability of the perturbative approach does not happen here, perhaps due to the nonperturbative nature of our approach.

\vspace{12pt}
\noindent {\bf Acknowledgments:}  We would like to thank an anonymous referee for interesting discussions and suggestions which helped to improve the quality of this work. 
 D.M.R is supported by Conselho Nacional de Desenvolvimento Cient\'\i fico e Tecnol\'ogico (CNPq) and Coordena\c c\~ao de Aperfeiçoamento de Pessoal de N\' \i vel Superior (Capes) (Brazilian Agencies), E.F.C. is partially supported by PROPGPEC-Col\'egio Pedro II, and H.B.-F. is partially supported by CNPq and Capes.

 \end{document}